\input{aipcheck}

\documentclass[
    ,final            
  ]
  {aipproc}

\layoutstyle{6x9}

\usepackage{amsmath}
\usepackage{graphicx}
\usepackage{grffile}
\usepackage{subfigure}

\newcommand{\ea}{\textit{et al}.\ }
\newcommand{\eq}[1]{(\ref{#1})}
\newcommand{\feynd}[1]{#1\kern-0.58em/}
\newcommand{\feynp}[1]{#1\kern-0.48em/}
\newcommand{\feynP}[1]{#1\kern-0.65em/}

\renewcommand{\P}{\mathcal{P}}
\newcommand{\p}{\partial}

\begin{document}

\title{Consistent interactions for high-spin baryons}

\classification{\mbox{11.10.Ef}, \mbox{11.15.-q}, \mbox{13.60.-r}, \mbox{13.75.-n}, \mbox{14.20.Gk}}
\keywords{Rarita-Schwinger fields, consistent interactions, hadron physics, hadron form factors}

\author{Tom Vrancx\footnote{Electronic address: {\ttfamily tom.vrancx@ugent.be}}, Lesley De Cruz, Jan Ryckebusch, and Pieter Vancraeyveld}{
  address={Department of Physics and Astronomy,\\
 Ghent University, Proeftuinstraat 86, B-9000 Gent, Belgium}
}


%
%

\begin{abstract}
Consistent interactions for off-shell fermion fields of arbitrary spin are constructed from the gauge-invariance requirement of the interaction Lagrangians. These interactions play a crucial role in the quantum hadrodynamical description of high-spin baryon resonances in hadronic processes.
We find that the power of the momentum dependence of a consistent interaction rises with the spin of the fermion field. This leads to unphysical structures in the energy dependence of the computed tree-level cross sections when the short-distance physics is cut off with standard hadronic form factors. A novel, spin-dependent hadronic form factor is proposed that suppresses the unphysical artifacts.
\end{abstract}

\maketitle

The Rarita-Schwinger (R-S) field $\psi_{\mu_1\ldots\mu_n}$ describes a relativistic spin-$(n+1/2)$ particle. By construction, the R-S field contains lower-spin components, which represent unphysical degrees of freedom. In the free R-S theory, these unphysical components are eliminated by the so-called R-S constraints \cite{Rarita:1941mf}. When considering interacting R-S fields, the lower-spin components only decouple if the R-S field is on its mass-shell (on-shell). For R-S fields that are off their mass shell (off-shell), the unphysical degrees of freedom do not decouple a priori. In this case the consistency of the interaction can only be assured if the interaction Lagrangian is invariant under the unconstrained spin-$(n+1/2)$ R-S gauge (uRS$_{n+1/2}$ gauge) \cite{Vrancx:2011}
\begin{align}
\psi_{\mu_1\ldots\mu_n} \rightarrow \psi_{\mu_1\ldots\mu_n} + \frac{i}{n(n-1)!}\sum_{P(\mu)}\p_{\mu_1}\chi_{\mu_2\ldots\mu_n}. \label{uRS_gauge}
\end{align}
With $\sum_{P(\mu)}$, one denotes a summation over all permutations of the $\mu_i$ indices and $\chi_{\mu_1\ldots\mu_{n-1}}$ represents an arbitrary, totally symmetric rank-$(n-1)$ tensor-spinor field.

Consistent interaction Lagrangians for the $(\phi\psi\psi^*_{\mu_1\ldots\mu_n})$- and $(A_\mu\psi\psi^*_{\mu_1\ldots\mu_n})$-theories are developed in Ref.~\cite{Vrancx:2011}, starting from the field $\Psi_{\mu_1\ldots\mu_n}$
\begin{align}
\Psi_{\mu_1\ldots\mu_n} = \sum_{P(\mu)}\sum_{k=0}^{n} \frac{i^n(-1)^k}{k!(n-k)!} \feynd{\p}^{\,k}\p_{\mu_{k+1}}\cdots\p_{\mu_n}\gamma^{\nu_{k+1}}\cdots\gamma^{\nu_n}\psi_{\mu_1\ldots\mu_k\nu_{k+1}\ldots\nu_n},\label{Psi}
\end{align}
which is explicitly invariant under the uRS$_{n+1/2}$ gauge \eq{uRS_gauge}. The fields $\phi$, $\psi$ and $A_\mu$ represent a spin-$0$, a spin-$1/2$ and a spin-$1$ field respectively. The gauge-invariant field $\Psi_{\mu_1\ldots\mu_n}$ of Eq.~\eq{Psi} gives rise to the following consistent interaction structure
\begin{align}
\hspace{-7.37pt}O^{n+{1/2}}_{(\mu_1\ldots\mu_n)\lambda_1\ldots\lambda_n}(p)P^{\lambda_1\ldots\lambda_n;\rho_1\ldots\rho_n}(p)O^{n+{1/2}}_{(\nu_1\ldots\nu_n)\rho_1\ldots\rho_n}(p) = p^{2n}\frac{\feynp{p} + m}{p^2 - m^2}\P^{n+{1/2}}_{\mu_1\ldots\mu_n;\nu_1\ldots\nu_n}(p), \label{interaction_structure-Psi}
\end{align}
where $P_{\mu_1\ldots\mu_n;\nu_1\ldots\nu_n}(p)$ represents the spin-$(n+1/2)$ R-S propagator that was derived in Ref.~\cite{Huang:2005js}, and $O^{n+{1/2}}_{(\mu_1\ldots\mu_n)\lambda_1\ldots\lambda_n}(p)$ is defined as $\Psi_{\mu_1\ldots\mu_n} = O^{n+{1/2}}_{(\mu_1\ldots\mu_n)\lambda_1\ldots\lambda_n}(\p)\psi^{\lambda_1\ldots\lambda_n}$. The simplified expression for the physical spin-$(n+1/2)$ projection operator $\P^{n+{1/2}}_{\mu_1\ldots\mu_n;\nu_1\ldots\nu_n}(p)$ is given in Ref.~\cite{Vrancx:2011}. 

\begin{figure}[t]
\centering
\includegraphics[scale=0.35]{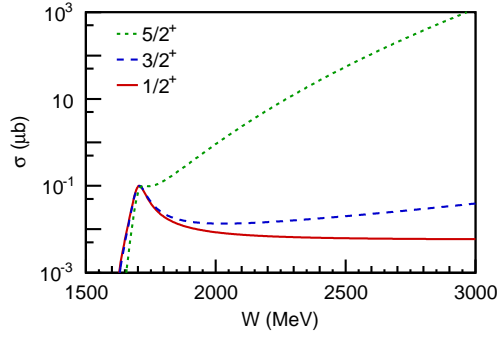}
\caption{The energy dependence of the $\gamma p \rightarrow N^* \rightarrow K^+\Lambda$ cross section. The $N^*$ is a mock resonance with $m_R = 1700$ MeV, $\Gamma_R = 50$ MeV and $J_R^P = 1/2^+$ (solid curve), $3/2^+$ (dashed curve), $5/2^+$ (dotted curve).} \label{spin_12-32-52}
\end{figure}

We wish to illustrate the use of consistent high-spin interactions with a detailed example: the construction of a tree-level amplitude for the reaction channel $\gamma p \rightarrow N^* \rightarrow K^+\Lambda$. The $N^*$ represents a nucleon resonance, which can be considered as the mediator of the mentioned reaction channel. The spin, mass and decay width of the $N^*$ are denoted as $J^P_R, m_R$ and $\Gamma_R$. It is worth stressing that the following discussion applies equally well to other hadronic processes that involve off-shell high-spin interactions. Fig.~\ref{spin_12-32-52} shows the computed energy dependence of the total cross section for three mock resonances with spins $J^P_R = 1/2^+, 3/2^+, 5/2^+$, $m_R = 1700$ MeV and $\Gamma_R = 50$ MeV.
The cross sections shown in Fig.~\ref{spin_12-32-52} display a high-energy behavior that is not acceptable from a physical point of view. In order to suppress this divergent high-energy behavior one usually introduces a hadronic form factor at the strong interaction vertex. The commonly used hadronic form factors are of the dipole form $F_d$ \cite{Pearce:1990uj} or the Gaussian form $F_G$ \cite{Corthals:2005ce}, i.e.\
\begin{align}
F_d(s;m_R,\Lambda_R) = \frac{\Lambda_R^4}{(s-m_R^2)^2+\Lambda_R^4}, \quad \textrm{or} \quad F_G(s;m_R,\Lambda_R) = \exp\left(-\frac{(s-m_R^2)^2}{\Lambda_R^4}\right), \label{standard_HFF}
\end{align}
where $s = W^2$ represents the squared invariant mass and $\Lambda_R$ the cut-off energy. 

For the remainder of this discussion, only one specific nucleon resonance will be considered, namely $N^* = N(1680)\;F_{15}$.
This resonance is an established $J^P_R = {5/2}^+$ nucleon resonance with a four-star rating in the Review of Particle Physics of the Particle Data Group (PDG) \cite{Nakamura:2010zzi}. It has a mass $m_R = 1685$ MeV and a decay width $\Gamma_R = 130$ MeV.
\begin{figure}[t]
\centering
\includegraphics[scale=0.35]{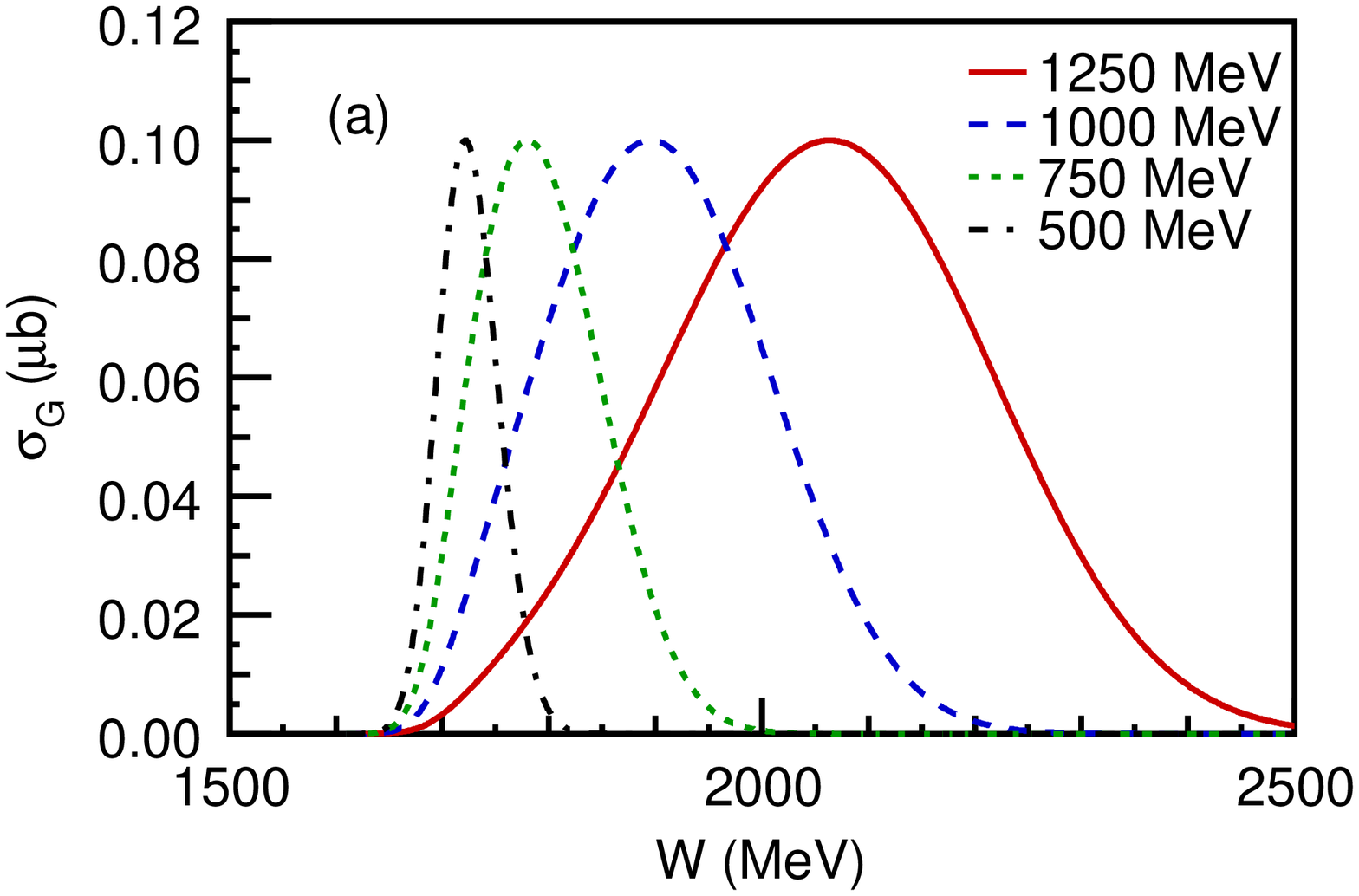}
\includegraphics[scale=0.35]{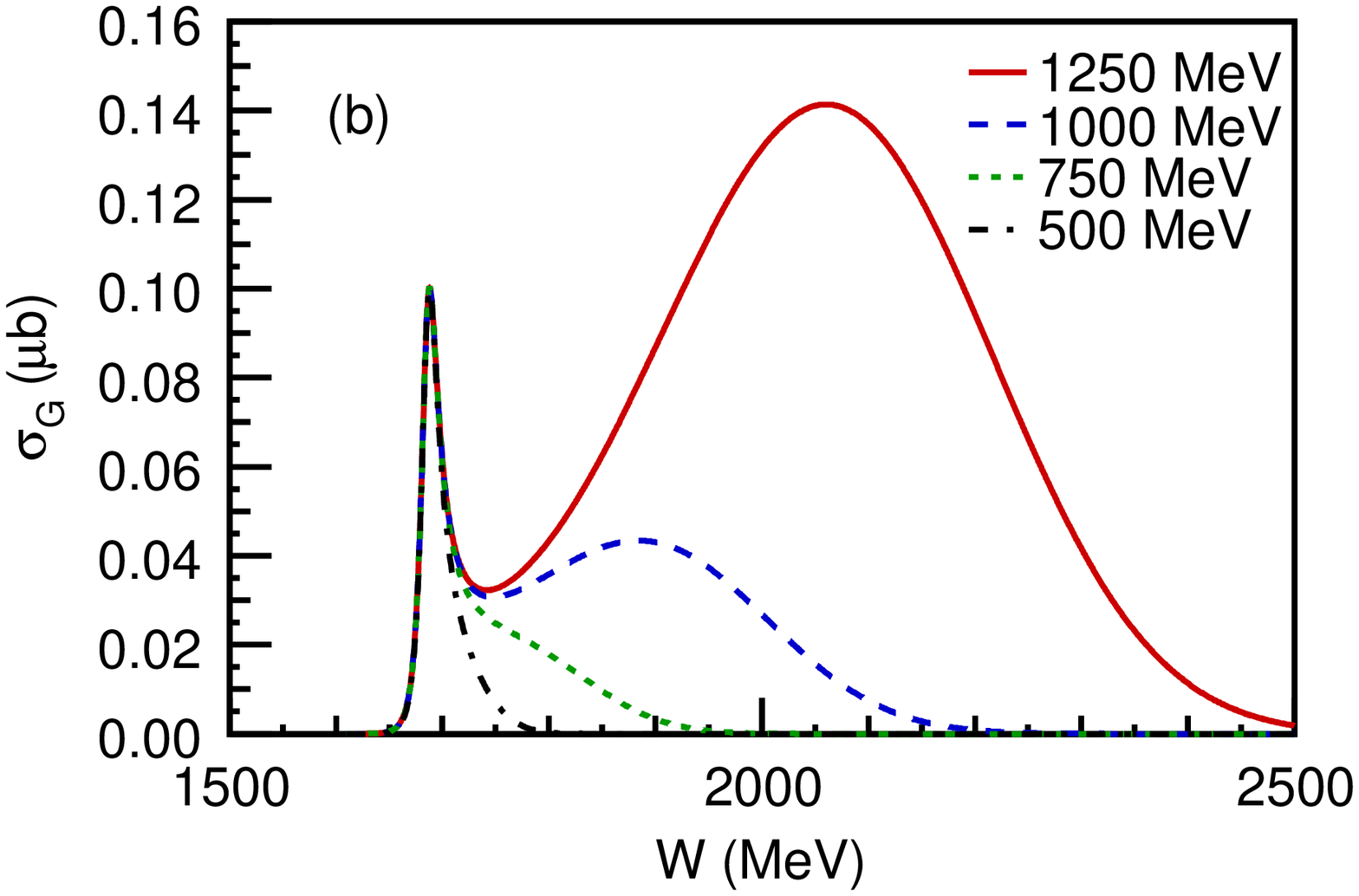}
\caption{The energy dependence of the $\gamma p \rightarrow N(1680)\;F_{15} \rightarrow K^+\Lambda$ cross section for various values of the Gaussian cut-off energy. In (a) the real decay width of the \mbox{$N(1680)\;F_{15}$} was used, i.e.\ $\Gamma_R = 130$ MeV. In (b) the decay width of the \mbox{$N(1680)\;F_{15}$} was set to $\Gamma_R' = 20$ MeV.} \label{spin-52_bumps-shift}
\end{figure}
In Fig.~\ref{spin-52_bumps-shift}(a) the computed energy dependence of the $\gamma p \rightarrow N(1680)\;F_{15} \rightarrow K^+\Lambda$ cross section is depicted for various values of the Gaussian cut-off energy. The structures that are observed, however, are artificial and do not correspond to the physical resonance peak of the \mbox{$N(1680)\;F_{15}$}. These artificial bumps arise from the combination of the divergent high-energy behavior of $\sigma$ and the fast decrease of $F_G$ with the energy. In addition, Fig.~\ref{spin-52_bumps-shift}(a) reveals that there is no indication of the actual resonance peak. This is explained by the relatively large decay width of the \mbox{$N(1680)\;F_{15}$} as compared to the fast increase of $\sigma$ with the energy. In Fig.~\ref{spin-52_bumps-shift}(b) the decay width of the \mbox{$N(1680)\;F_{15}$} was artificially lowered to $\Gamma'_R = 20$ MeV. This figure confirms both previous statements, namely the artificial nature of the observed structures and the absence of the actual resonance peaks. It also shown that for relatively small decay widths the artificial bump can be suppressed by lowering the cut-off energy. However, most of the nucleon resonances listed by PDG (if not all) have a relatively large decay width \cite{Nakamura:2010zzi}, as is the case for the \mbox{$N(1680)\;F_{15}$}. For these resonances lowering the cut-off energy results in a mere shift of the artificial bump towards the threshold energy. It is important to stress that with the dipole form factor $F_d$, defined in Eq.~\eq{standard_HFF}, things get even worse. Indeed, the decrease of the dipole form factor with the energy is not sufficient to compensate for the divergent high-energy behavior of $\sigma_d$.

\begin{figure}[t]
\centering
\includegraphics[scale=0.35]{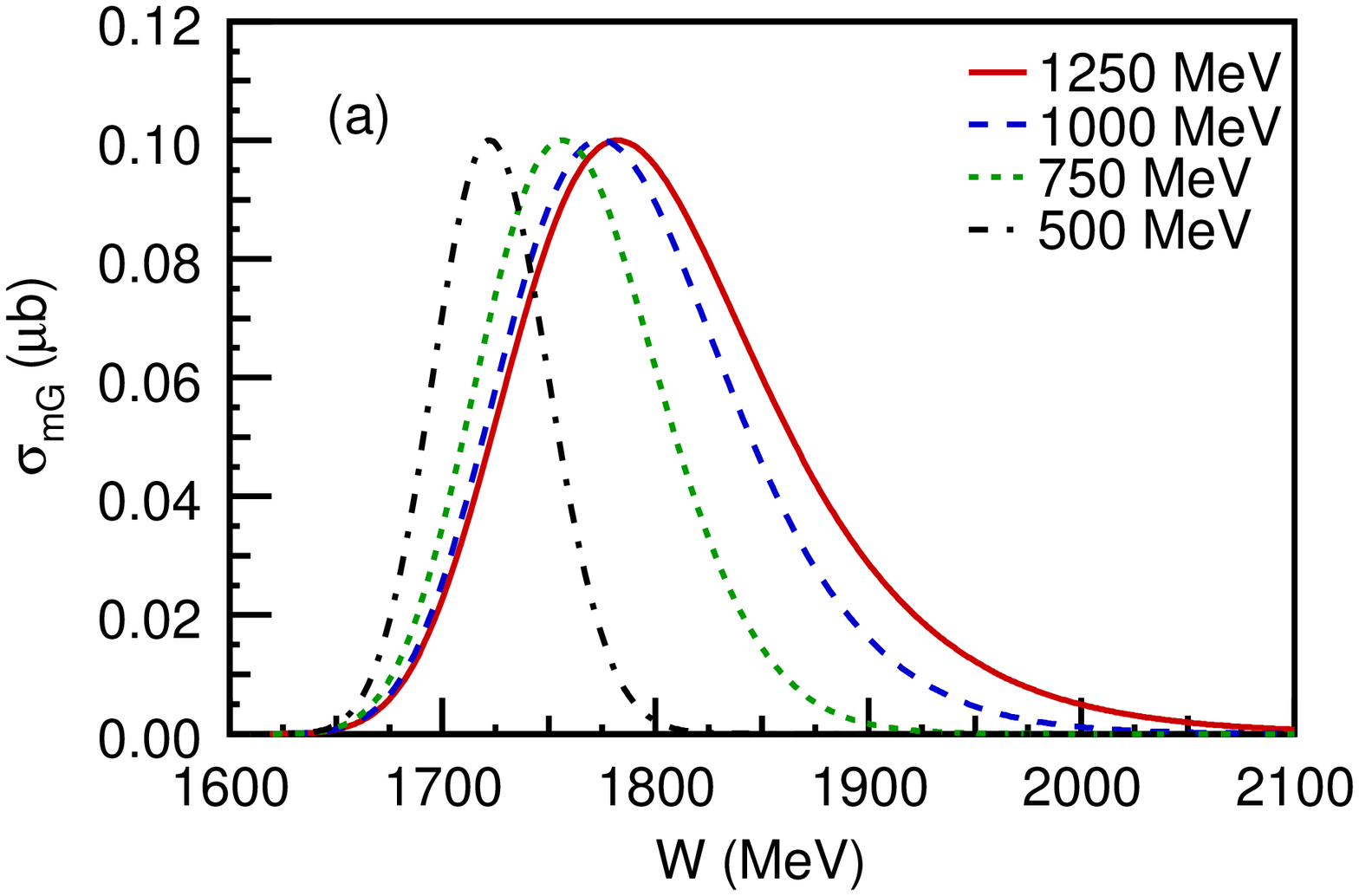}
\includegraphics[scale=0.35]{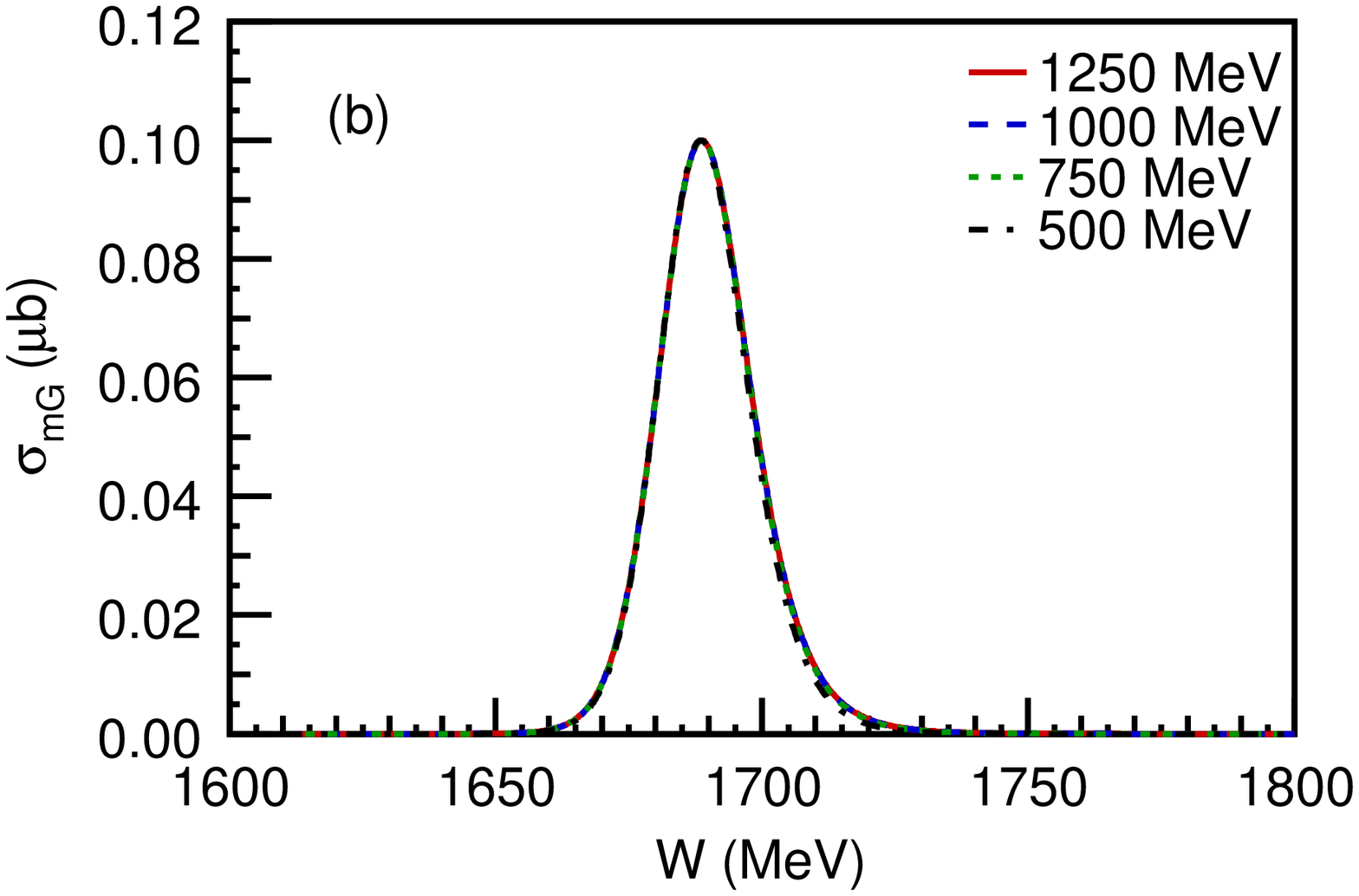}
\caption{The energy dependence of the $\gamma p \rightarrow N(1680)\;F_{15} \rightarrow K^+\Lambda$ cross section for various values of the cut-off energy of the multidipole-Gauss form factor. In (a) the real decay width of the \mbox{$N(1680)\;F_{15}$} was used, i.e.\ $\Gamma_R = 130$ MeV. In (b) the decay width of the \mbox{$N(1680)\;F_{15}$} was set to $\Gamma_R' = 20$ MeV.}\label{spin-52_mG-HFF-C}
\end{figure}

The above-mentioned issues with regard to the energy position of the peaks can be resolved by introducing the so-called multidipole-Gauss form factor \cite{Vrancx:2011}
\begin{align}
F_{mG}(s;m_R,\Lambda_R,\Gamma_R,J_R) = \left(\frac{m_R^2\widetilde{\Gamma}^2_R(J_R)}{(s-m_R^2)^2+m_R^2\widetilde{\Gamma}^2_R(J_R)}\right)^{J_R-\frac{1}{2}}\exp\left(-\frac{(s-m_R^2)^2}{\Lambda_R^4}\right),\label{mG_HFF}
\end{align}
where $\widetilde{\Gamma}_R(J_R)$ is the spin-dependent modified decay width
\begin{align}
\widetilde{\Gamma}_R(J_R) = \frac{\Gamma_R}{\sqrt{2^{\frac{1}{2J_R}}-1}}. \label{modified_decay_width}
\end{align}
The dipole part of $F_{mG}$ raises the multiplicity of the propagator pole. In this way the divergent high-energy behavior of $\sigma$, which is shown in Fig.~\ref{spin_12-32-52}, is regulated. As a result, $F_{mG}$ removes the artificial bump in the computed energy dependence of the cross section, while restoring the resonance peak of the decaying particle. Fig.~\ref{spin-52_mG-HFF-C}(a) shows the computed energy dependence of the $\gamma p \rightarrow N(1680)\;F_{15} \rightarrow K^+\Lambda$ cross section for various values of the cut-off energy of the multidipole-Gauss form factor. Here it appears that the observed mass (i.e.\ the energy that corresponds to the peak value of the structure) and decay width (i.e.\ the full width at half maximum of the structure) of the \mbox{$N(1680)\;F_{15}$} do not correspond to their physical values, but instead are function of the cut-off energy. However, this is just a threshold effect. This is confirmed by Fig.~\ref{spin-52_mG-HFF-C}(b) in which the decay width of the \mbox{$N(1680)\;F_{15}$} was again lowered to $\Gamma'_R = 20$ MeV. Here, the threshold effects are largely reduced and the deviation of the observed mass and decay width from the physical values amounts to respectively $0.22$\% and $0.75$\%.

\begin{theacknowledgments}
This work was supported by the Fund for Scientific Research Flanders and the Research Council of Ghent University.
\end{theacknowledgments}

\bibliographystyle{aipproc}

\end{document}